\documentclass{elsart}
\journal{Physics Letter B}
\usepackage{epsfig}
\usepackage{bm}
\usepackage{graphicx}

\newcommand{\ra}{\rightarrow}
\newcommand{\ccs}{c\bar{c}s}

\def\ccd{c\bar{c}d}
\def\btoccd{b \to \ccd}

\newcommand{\btoccs}{b \ra \ccs}

\def\bz{B^0}
\def\bzb{\overline{B}{}^0}
\def\fCP{f_{CP}}
\newcommand{\bp}{B^+}
\newcommand{\bminus}{B^-}

\newcommand{\ks}{K_{\rm S}^0}

\newcommand{\dsm}{D_{\rm SM}}

\def\cals{\mathcal{S}}
\def\cala{\mathcal{A}}

\def\dm{\Delta m_d}
\def\dmd{\dm}
\def\taubz{\tau_\bz}
\def\taub{\taubz}
\def\dsp{D^{*+}}
\def\dsm{D^{*-}}
\def\fq{\ensuremath{q}}
\def\cal{\mathcal}

\def\bz{{B^0}}
\def\bzb{{\overline{B}{}^0}}
\def\kl{K_L^0}
\def\dE{{\Delta E}}
\def\de{{\Delta E}}
\def\mb{{M_{\rm bc}}}
\def\Dt{\Delta t}

\def\fol{f_{\rm ol}}
\def\fsig{f_{\rm sig}}
\newcommand{\sinbb}{{\sin2\phi_1}}

\newcommand{\ftag}{f_{\rm tag}}

\newcommand*{\dwl}{\ensuremath{{\Delta w_l}}}

\def\bztodspdsm{\bz \to \dsp \dsm}
\def\dzero{D^0}
\def\dzerob{\overline{D}{}^0}
\def\dzerobstar{\overline{D}{}^{*0}}
\def\call{\mathcal L}
\def\calb{\mathcal B}

\newcommand{\ttr}{\theta_{\rm tr}}
\newcommand{\tone}{\theta_{1}}
\newcommand{\phitr}{\phi_{\rm tr}}
\def\Hzero{H_0}
\def\Hpara{H_\|}
\def\Hperp{H_\perp}
\def\calh{\mathcal H}

\newcommand{\rzero}{R_{0}}

\newcommand{\rperp}{R_{\perp}}

\def\adsds{\cala}
\def\sdsds{\cals}

\newcommand{\sdsdscenter}{-0.75}
\newcommand{\sdsdsstat}  { 0.56}
\newcommand{\sdsdssyst}  { 0.12}
\newcommand{\sdsdsresult}{\sdsdscenter\pm\sdsdsstat{\rm(stat)}\pm\sdsdssyst{\rm(syst)}}
\newcommand{\adsdscenter}{-0.26}
\newcommand{\adsdsstat}  { 0.26}
\newcommand{\adsdssyst}  { 0.06}
\newcommand{\adsdsresult}{\adsdscenter\pm\adsdsstat{\rm(stat)}\pm\adsdssyst{\rm(syst)}}
\newcommand{\rperpcenter}{0.19}
\newcommand{\rperpstat}  {0.08}
\newcommand{\rperpsyst}  {0.01}
\newcommand{\rperpresult}{\rperpcenter\pm\rperpstat{\rm(stat)}\pm\rperpsyst{\rm(syst)}}
\newcommand{\rzerocenter}{0.57}
\newcommand{\rzerostat}  {0.08}
\newcommand{\rzerosyst}  {0.02}
\newcommand{\rzeroresult}{\rzerocenter\pm\rzerostat{\rm(stat)}\pm\rzerosyst{\rm(syst)}}
\newcommand{\brdsdscenter}{0.81}
\newcommand{\brdsdsstat}  {0.08}
\newcommand{\brdsdssyst}  {0.11}
\newcommand{\brdsdsresult}{[\brdsdscenter\pm\brdsdsstat{\rm(stat)}\pm\brdsdssyst{\rm(syst)}]\times10^{-3}}

\begin{document}

%%% Comment out the following two lines 
%%% for PLB submission
%\flushright{Version 0.20}
%\vskip 3cm
%%%

\begin{frontmatter}

% <<< for preprint >>>
\vspace*{-3\baselineskip}
\begin{flushleft}
 \resizebox{!}{3cm}{\includegraphics{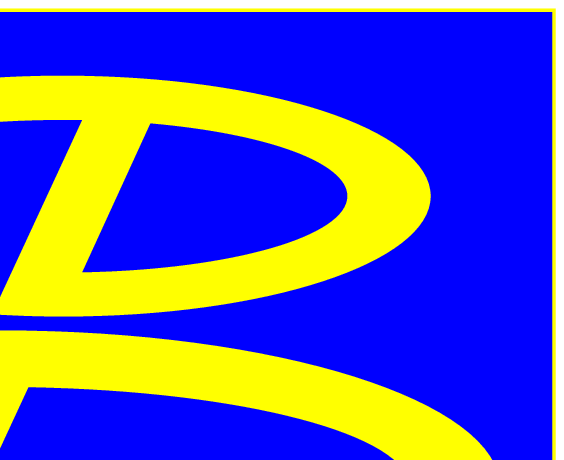}}
\end{flushleft}
\vspace*{-3cm}
\begin{flushright}
 Belle Preprint 2004-41\\
 KEK Preprint 2004-82
\end{flushright}
\vspace*{2cm}

\title{\boldmath Branching Fraction, Polarization and $CP$-Violating Asymmetries
in $\bz \to \dsp \dsm$ Decays}

%\author{Belle Collaboration}
%\input{author}
%%% Paper:    B -> D*D* BF/polarization/CPV
%%% Journal:  Physics Letters B
%%% Contacts: H. Miyake (miyake@champ.hep.sci.osaka-u.ac.jp)
%%%           M. Hazumi (masashi.hazumi@kek.jp)
%%% Non-responding authors or those who said NO are commented out.
%%% ====================================================================
%%% Click the RELOAD button on your web browser to see the updated file.
%%% ====================================================================
%%% Use \input{author} to insert this material into your latex file.
\collab{Belle Collaboration}
  \author[Osaka]{H.~Miyake}, % Osaka
  \author[KEK]{M.~Hazumi}, % KEK
  \author[KEK]{K.~Abe}, % KEK
  \author[TohokuGakuin]{K.~Abe}, % TohokuGakuin
% \author[TIT]{N.~Abe}, % TIT
% \author[KEK]{I.~Adachi}, % KEK
  \author[Tokyo]{H.~Aihara}, % Tokyo
% \author[Nagoya]{M.~Akatsu}, % Nagoya
  \author[Tsukuba]{Y.~Asano}, % Tsukuba
% \author[Toyama]{T.~Aso}, % Toyama
  \author[BINP]{V.~Aulchenko}, % BINP
  \author[ITEP]{T.~Aushev}, % ITEP
  \author[Tata]{T.~Aziz}, % Tata
  \author[Cincinnati]{S.~Bahinipati}, % Cincinnati
% ...
% Ljubljana

% \author[Lausanne]{D.~Z\"urcher}, % Lausanne

%%%\address[Aomori]{Aomori University, Aomori, Japan}
\address[BINP]{Budker Institute of Nuclear Physics, Novosibirsk, Russia}
% ...
\address[Cincinnati]{University of Cincinnati, Cincinnati, OH, USA}
%%%\address[Frankfurt]{University of Frankfurt, Frankfurt, Germany}
% ...
\address[KEK]{High Energy Accelerator Research Organization (KEK), Tsukuba, Japan}
% ...
\address[ITEP]{Institute for Theoretical and Experimental Physics, Moscow, Russia}
% ...
\address[Osaka]{Osaka University, Osaka, Japan}
% ...
\address[Tata]{Tata Institute of Fundamental Research, Bombay, India}
% ...
\address[TohokuGakuin]{Tohoku Gakuin University, Tagajo, Japan}
% ...
\address[Tokyo]{Department of Physics, University of Tokyo, Tokyo, Japan}
% ...
\address[Tsukuba]{University of Tsukuba, Tsukuba, Japan}
%%%\address[Utkal]{Utkal University, Bhubaneswer, India}
% ...
\thanks[NovaGorica]{on leave from Nova Gorica Polytechnic, Nova Gorica, Slovenia}
%
%\date{\today}

\begin{abstract}

We present measurements of the branching fraction, 
the polarization parameters
and $CP$-violating asymmetries in $\bz \to \dsp \dsm$ decays
using a 140 fb$^{-1}$ data sample collected at the
$\Upsilon(4S)$ resonance with the Belle detector at
the KEKB energy-asymmetric $e^+e^-$ collider. 
We obtain 
$\calb(\bztodspdsm) = \brdsdsresult,~
\rperp = \rperpresult,~
\rzero = \rzeroresult,~
\sdsds = \sdsdsresult$ and $
\adsds = \adsdsresult.$
Consistency with Standard Model expectations is also discussed.
\end{abstract}


\end{frontmatter}

%%%%%%%%%%%%%%%%%%%%%%%%%%%%
\section{Introduction}               % Introduction goes below.
\label{sec:intro}

In the Standard Model (SM),
$CP$ violation arises from an irreducible complex phase,
the Kobayashi-Maskawa (KM) phase~\cite{bib:KM},
in the weak-interaction quark-mixing matrix.
In particular, the SM predicts $CP$ asymmetries in 
the time-dependent rates for $\bz$ and
$\bzb$ decays to a common $CP$ eigenstate $\fCP$~\cite{bib:sanda}. 
Recent measurements of the $CP$-violation parameter $\sin2\phi_1$
by the Belle~\cite{bib:CP1_Belle,bib:CP140PRD_Belle}
and BaBar~\cite{bib:CP1_BaBar} collaborations established $CP$ violation
in $\bz \to J/\psi \ks$ and related decay modes~\cite{bib:CC},
which are governed by the $b \to c\overline{c}s$ transition,
at a level consistent with KM expectations.
Here $\phi_1$ is one of the three interior angles of 
the Unitarity Triangle~\cite{bib:CP1_Belle,bib:CP140PRD_Belle}.

Despite this success, many tests remain before it can be concluded
that the KM phase is the only source of $CP$ violation.
The $\bz\to \dsp\dsm$ decay, which is dominated by the $\btoccd$ transition,
provides an additional test of the SM.
Within the SM, 
measurements of $CP$ violation in this mode
should yield the $\sin 2\phi_1$ value to a good approximation
if the contribution from the penguin diagram is neglected. 
The correction from the penguin diagram is expected to
be small~\cite{bib:Pham1999}.
Thus, a significant deviation in the time-dependent $CP$ asymmetry in these 
modes from what is observed 
in $\btoccs$ decays would be evidence for a new $CP$-violating phase.

In the decay chain $\Upsilon(4S)\to \bz\bzb \to f_{CP}f_{\rm tag}$,
where one of the $B$ mesons decays at time $t_{CP}$ to a final state $f_{CP}$ 
and the other decays at time $t_{\rm tag}$ to a final state  
$f_{\rm tag}$ that distinguishes between $B^0$ and $\bzb$, 
the decay rate has a time dependence
given by~\cite{bib:sanda}
\begin{equation}
\label{eq:psig}
{\mathcal P}(\Delta{t}) = 
\frac{e^{-|\Delta{t}|/{\taubz}}}{4{\taubz}}
\biggl\{1 + \fq
\Bigl[ \cals\sin(\dmd\Delta{t})
   + \cala\cos(\dmd\Delta{t})
\Bigr]
\biggr\}.
\end{equation}
Here $\cals$ and $\cala$ are $CP$-violation parameters, 
$\taubz$ is the $B^0$ lifetime, $\dmd$ is the mass difference 
between the two $B^0$ mass
eigenstates, $\Delta{t}$ = $t_{CP}$ $-$ $t_{\rm tag}$, and
$\fq$ = +1 ($-1$) when the tagging $B$ meson
is a $B^0$ 
($\bzb$).
The parameter 
$\cals$ corresponds to the mixing-induced $CP$ violation and is 
related to $\sin 2\phi_1$, while $\cala$ represents direct $CP$ violation that
normally arises from the interference between tree and penguin diagrams. 

In $\bz\to \dsp\dsm$ decays the final state
$D^*$ mesons may be in a state of $s$-, $p$- or $d$-wave relative 
orbital angular momentum. 
Since $s$- and $d$-waves are even under the $CP$ transformation
while the $p$-wave is odd,
the $CP$-violation parameters in Eq.~(\ref{eq:psig}) are diluted.
In order to determine the dilution, one needs to measure
the $CP$-odd fraction. This can be accomplished
with a time-integrated angular analysis.
The BaBar collaboration has measured the polarization and $CP$
asymmetries~\cite{bib:DstarDstar_BaBar}, 
and find the $CP$-odd contribution to be small,
consistent with theoretical expectations~\cite{bib:Pham1999}.
The $CP$ asymmetries are found to differ slightly from the expectation
that neglects the contribution from the penguin diagram.

In this Letter we report measurements of
the branching fraction, the polarization parameters and $CP$ asymmetries
in $\bz \to \dsp \dsm$ decays
based on a 140 fb$^{-1}$ data sample,
which corresponds to 152 million $B\overline{B}$ pairs.
At the KEKB energy-asymmetric 
$e^+e^-$ (3.5 on 8.0~GeV) collider~\cite{bib:KEKB},
the $\Upsilon(4S)$ is produced
with a Lorentz boost of $\beta\gamma=0.425$
antiparallel to the positron beamline ($z$).
Since the $B^0$ and $\bzb$ mesons are approximately at 
rest in the $\Upsilon(4S)$ center-of-mass system (cms),
$\Delta t$ can be determined from the displacement in $z$ 
between the $f_{CP}$ and $f_{\rm tag}$ decay vertices:
$\Delta t \simeq (z_{CP} - z_{\rm tag})/(\beta\gamma c)
 \equiv \Delta z/(\beta\gamma c)$.

The Belle detector~\cite{bib:Belle} is a large-solid-angle spectrometer
that includes a three-layer silicon vertex detector (SVD),
a 50-layer central drift chamber (CDC),
an array of aerogel threshold Cherenkov counters (ACC),
time-of-flight (TOF) scintillation counters,
and an electromagnetic calorimeter comprised of CsI(Tl) crystals (ECL)
located inside a superconducting solenoid coil
that provides a 1.5~T magnetic field.
An iron flux-return located outside of the coil is instrumented
with resistive plate chambers to detect $\kl$ mesons and 
to identify muons (KLM).

%%%%%%%%%%%%%%%%%%%%%%%%%%%%%%%%%%%%%%%%%%%%
\section{Event Selection}
\label{sec:evsel}

We reconstruct $\bztodspdsm$ decays in the following $D^*$ final states;
$(\dzero\pi^+,\\ \dzerob\pi^-)$,
$(\dzero\pi^+, D^-\pi^0)$ and
$(D^+\pi^0, \dzerob\pi^-)$.
For the $\dzero$ decays we use
$\dzero \to K^-\pi^+$, $K^-\pi^+\pi^0$, $K^-\pi^+\pi^+\pi^-$, $K^+K^-$,
$\ks \pi^+\pi^-$ and $\ks \pi^+\pi^-\pi^0$.
For the $D^+$ decays we use 
$D^+ \to \ks\pi^+$, $\ks\pi^+\pi^0$, $\ks K^+$, $K^-\pi^+\pi^+$
and $K^-K^+\pi^+$.
We allow all combinations of $D$ decays except
for cases where both $D$ decays include
neutral kaons in the final state.

Charged tracks from $D$ meson decays are required to be
consistent with originating from
the interaction point (IP).
Charged kaons are separated from pions according to
the likelihood ratio $P_{K/\pi} \equiv \call(K)/[\call(K)+\call(\pi)]$,
where the likelihood function $\call$ is
based on the combined information from
the ACC, CDC $dE/dx$ and TOF measurements.
We require $P_{K/\pi} > 0.1~(0.2)$ for kaons in
2-prong (4-prong) $D$ meson decays.
The kaon identification efficiency is $96\%$, and $13\%$ of pions are
misidentified as kaons.
Candidate charged pions are required to satisfy $P_{K/\pi} < 0.9$,
which provides a pion selection efficiency of $91\%$ with a kaon 
misidentification probability of $3\%$.
Neutral pions are formed from two photons with
invariant masses above 119 MeV/$c^2$ and below 146 MeV/$c^2$.
To reduce the background from low-energy photons,
we require $E_\gamma > 0.03$ GeV for each photon
and $p_{\pi^0} > 0.1$ GeV/$c$,
where $E_\gamma$ and $p_{\pi^0}$ are the photon energy and
the $\pi^0$ momentum in the laboratory frame, respectively.
Candidate $\ks \to \pi^+\pi^-$ decays are reconstructed
from oppositely charged track pairs that have invariant
masses within 15 MeV/$c^2$ of the nominal $\ks$ mass.
A reconstructed $\ks$ is required to have a displaced vertex and a
flight direction consistent with that of a $\ks$ originating from
the IP.

\begin{figure}
\begin{center}
\resizebox{0.4\textwidth}{!}{\includegraphics{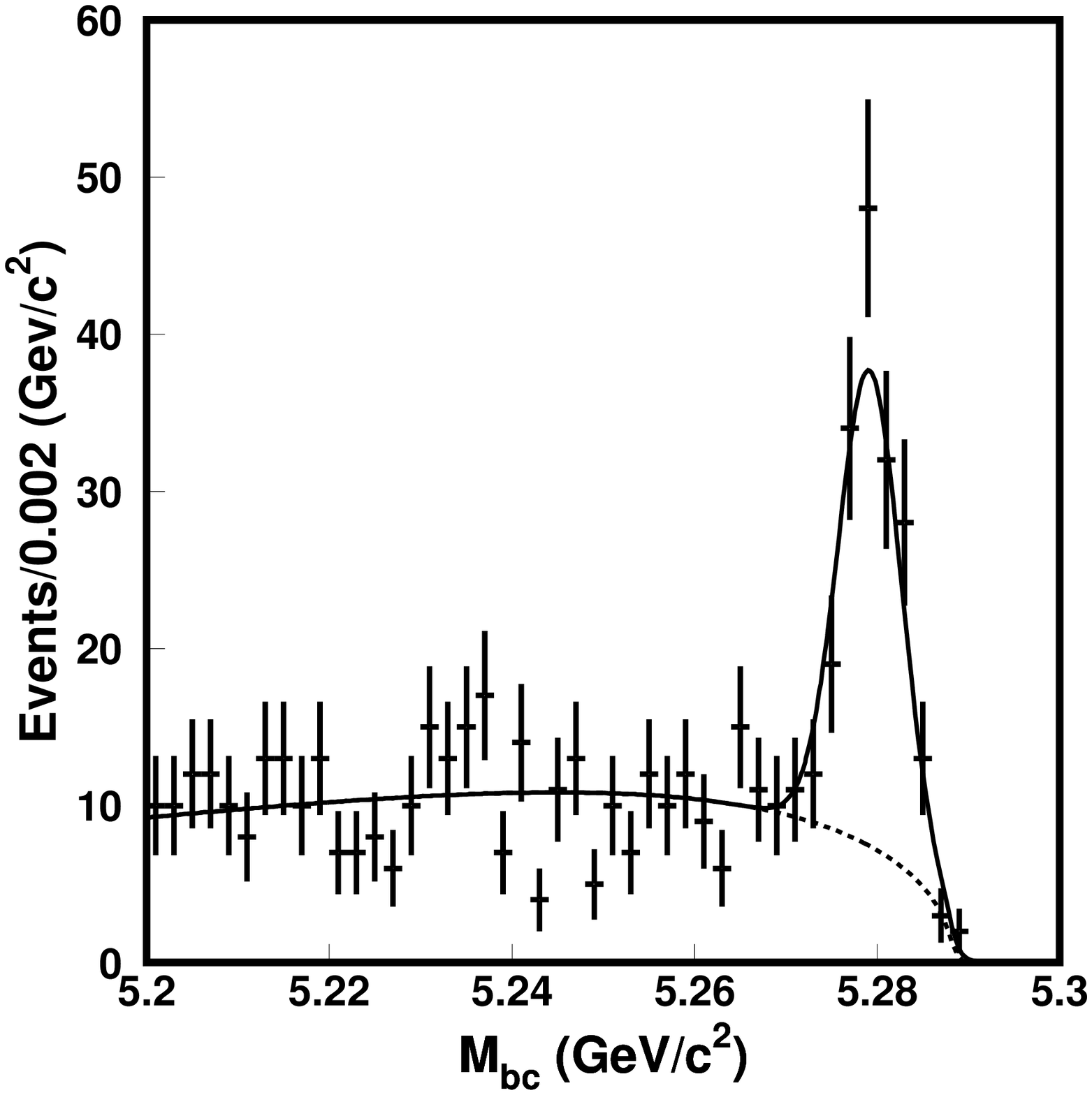}}
\resizebox{0.4\textwidth}{!}{\includegraphics{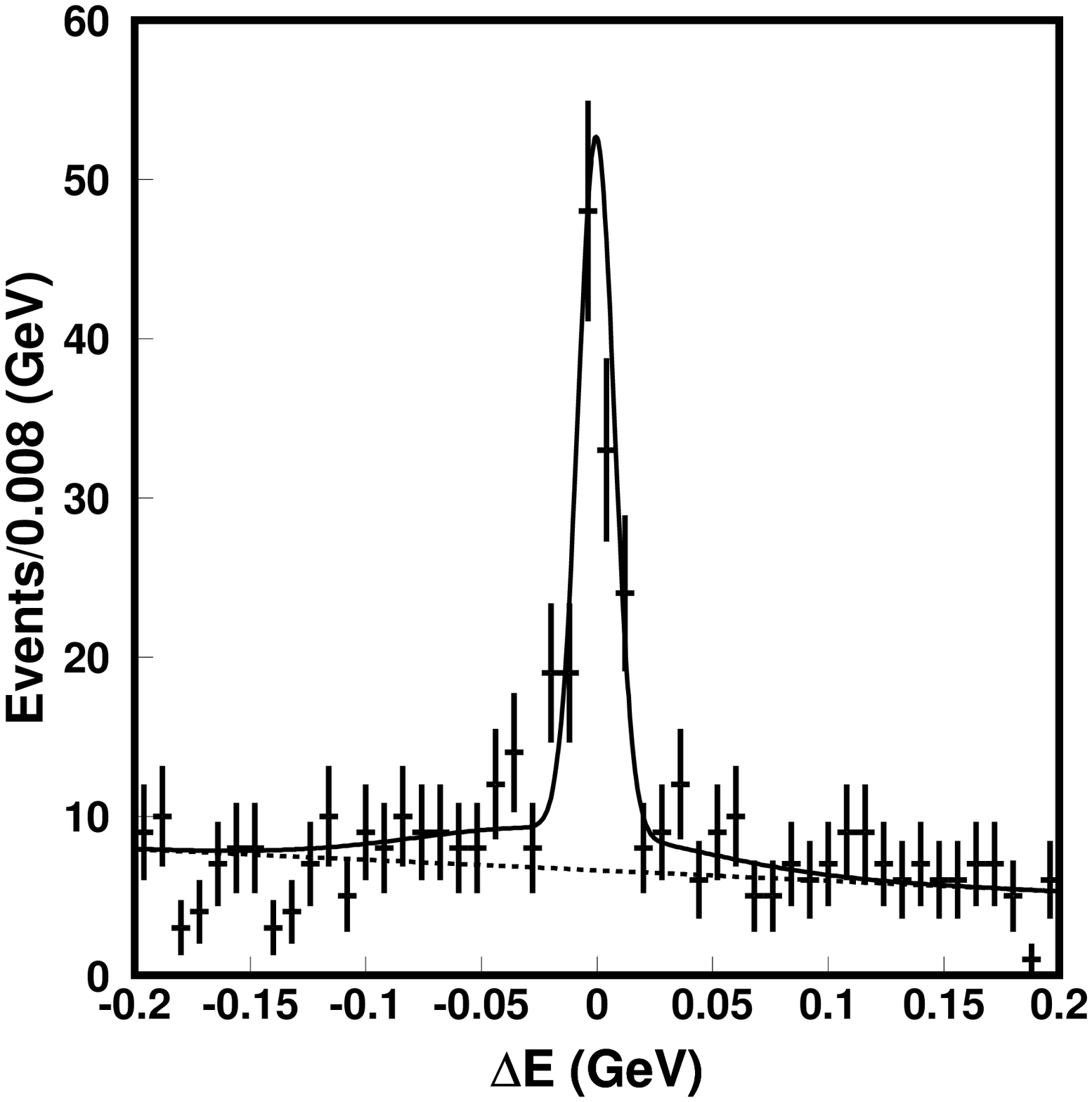}}
\end{center}
\caption{ (Left) $\mb$ and (right) $\dE$ distributions for
$\bztodspdsm$ candidates within the $\dE$ ($\mb$) signal region.
Solid curves show the fit to signal plus background distributions,
and dashed curves show the background contributions that comprise
$\bp\bminus$, $\bz\bzb$ and continuum events.
}\label{fig:mb}
\end{figure}

Candidate $D$ mesons are reconstructed from
the selected kaons and pions,
and are required to have invariant masses
within 6$\sigma$ (3$\sigma$) of the $D$ meson mass
for 2-prong (3- or 4-prong) decays, where $\sigma$ is the mass resolution
that ranges from $5$ to $10$ MeV/$c^2$.
In this selection $\sigma$ is obtained by fitting 
the Monte Carlo (MC) simulated $D$ meson mass.
These $\dzero$ ($D^+$) candidates are then combined with
$\pi^+$ ($\pi^0$) to form $\dsp$ candidates, where
the IP and pion identification requirements are not used
to select $\pi^+$ candidates.
The mass difference between $\dsp$ and $\dzero$ ($D^+$) is required to
be within 3.00 (2.25) MeV/$c^2$ of the nominal mass difference.
We identify $B$ meson decays using the
energy difference $\dE\equiv E_B^{\rm cms}-E_{\rm beam}^{\rm cms}$ and
the beam-energy constrained mass $\mb\equiv\sqrt{(E_{\rm beam}^{\rm cms})^2-
(p_B^{\rm cms})^2}$, where $E_{\rm beam}^{\rm cms}$ is
the beam energy in the cms, and
$E_B^{\rm cms}$ and $p_B^{\rm cms}$ are the cms energy and momentum,
respectively, of the reconstructed $B$ candidate.
The $B$ meson signal region is defined as 
$|\dE|<0.04$ GeV and $\mb$ within 3$\sigma$ of the $B$ meson mass,
where $\sigma$ is $3.5$ MeV/$c^2$.
In order to suppress background from the $e^+e^- \rightarrow 
u\overline{u},~d\overline{d},~s\overline{s}$, or $c\overline{c}$
continuum, we require $H_2/H_0 < 0.4$, where $H_2$ ($H_0$) is the 
second (zeroth) Fox-Wolfram moment~\cite{bib:FW}.
After applying this requirement, we find that the contributions
to the background from $\bp\bminus$, $\bz\bzb$ and continuum are
approximately equal.
Figure~\ref{fig:mb} shows the $\mb$ and $\dE$ distributions for the
$\bztodspdsm$ candidates that are in the $\dE$ and $\mb$ signal regions,
respectively.
%%%%%%%%%%
%%%%%%%%%%
In the $\mb$ and $\dE$ signal regions there are 194 events.

%%%%%%%%%%%%%%%%%%%%%%%%%%%%%%%
\section{Branching Fraction}
\label{sec:br}

To determine the signal yield, we perform a two-dimensional
maximum likelihood fit to the $\mb$-$\dE$ distribution
($5.2$ GeV/$c^2 < \mb < 5.3$ GeV/$c^2$ and $|\dE| < 0.2$ GeV).
We use a Gaussian signal distribution plus the ARGUS
background function~\cite{bib:ARGUS} for the $\mb$ distribution.
The signal shape parameters are determined from MC.
The background parameters
are obtained simultaneously in the fit to data.
The $\dE$ distribution is modeled by
a double Gaussian signal function plus a linear background function.
We obtain shape parameters separately for candidates
with and without $\dsp \to D^+\pi^0$ decays
to account for small differences between the two cases.

The fit yields $130 \pm 13$ signal events, where
20\% include $\dsp \to D^+\pi^0$ decays.
To obtain the branching fraction $\calb(\bztodspdsm)$,
we use the reconstruction efficiency and
the known branching fraction for each subdecay mode.
We obtain an effective efficiency of $[1.06\pm0.08] \times 10^{-3}$ from 
the sum of the products of MC reconstruction efficiencies
and branching fractions for each of the subdecays. 
Small corrections are applied to the 
reconstruction efficiencies for charged tracks, neutral pions and
$\ks$ mesons to account for differences between data and MC.

We obtain
\begin{equation}
\calb(\bztodspdsm) = 
\brdsdsresult,
\end{equation}
where the first error is statistical and the second is systematic.
The result is consistent with the present world-average 
value~\cite{bib:PDG2003}.

The dominant sources of the systematic error are
uncertainties in the tracking efficiency (11\%) and
in the subdecay branching fractions (7\%).
Other sources are uncertainties 
in the fit parameters and methods (1\%),
in the reconstruction efficiencies of $\pi^0~(2\%)$ and $\ks~(1\%)$,
particle identification $(1\%)$, polarization parameters $(2\%)$,
the number of $B$ mesons $(1\%)$, and MC statistics $(1\%)$,
%where each value in parentheses is the total contribution in percent.
where each value in parentheses is the total contribution.

%%%%%%%%%%%%%%%%%%%%%%%%%%%%%%%%%%%%
\section{Polarization} 
\label{sec:pol}

The time-dependent $CP$ analysis requires knowledge of the $CP$-odd fraction.
To obtain the $CP$-odd fraction without bias, 
we must take into account the efficiency difference between the two
$CP$-even components. Therefore, we perform a time-integrated
two-dimensional angular analysis to obtain the fraction of each
polarization component.
We use the transversity basis~\cite{bib:Dunietz:1990cj}
where three
angles $\tone$, $\ttr$ and $\phitr$ are defined in Fig. 2.
The angle $\tone$ is the angle between the momentum of the slow pion
from the $\dsm$ in the $\dsm$ rest frame and the direction
opposite to $B$ momentum in the $\dsm$ rest frame.
The angle $\ttr$ is the polar angle between the
normal to the $\dsm$ decay plane and the direction of flight
of the slow pion from the $\dsp$ in the $\dsp$ rest frame.
The angle $\phitr$ is the corresponding azimuthal angle,
where $\phitr=0$ is the direction antiparallel to the $\dsm$ flight direction.
%%%%%%%%%%
\begin{figure}
\begin{center}
\resizebox{0.4\textwidth}{!}{\includegraphics{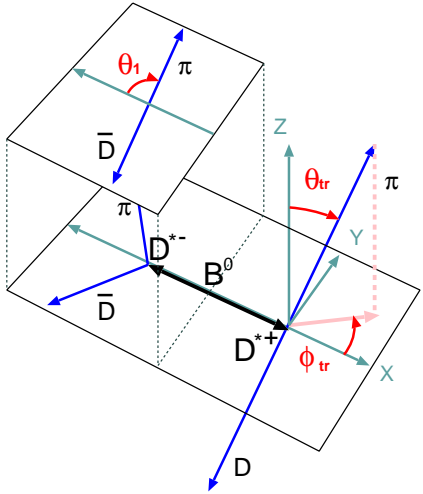}}
\end{center}
\caption{Definition of the angles in the transversity basis.
  Angle $\ttr$ and $\phitr$ are defined in the $\dsp$ rest frame
 (the lower plane),
  while $\tone$ is defined in the $\dsm$ rest frame (the upper plane).
}
\label{fig:tra}
\end{figure}
%%%%%%%%%%%
Integrating over time and the angle $\phitr$, the
two-dimensional differential decay rate is
\begin{equation}
\frac{1}{\Gamma}\frac{d^2\Gamma}{d\cos\ttr d\cos\tone}
 = \frac{9}{16}\sum_{i=0,\|,\perp}{R_iH_i(\cos\ttr,\cos\tone)},
\label{eq:root_angular}
\end{equation}
where $i=0,\|,$ or $\perp$ denotes longitudinal, transverse parallel,
or transverse perpendicular components, $R_i$ is its fraction
that satisfies
\begin{equation}
R_0+R_{\|}+R_{\perp}=1,
\end{equation}
and $H_i$ is its angular distribution defined as
\begin{eqnarray}
\Hzero(\cos\ttr,\cos\tone) &= 2 \sin^2\ttr \cos^2\tone,&\nonumber \\
\Hpara(\cos\ttr,\cos\tone) &= \phantom1\sin^2\ttr \sin^2\tone,&\\
\Hperp(\cos\ttr,\cos\tone) &= 2 \cos^2\ttr \sin^2\tone.&\nonumber
\end{eqnarray}
The fraction $R_\perp$ corresponds to the $CP$-odd fraction.

Eq.~(\ref{eq:root_angular}) is affected by the detector
efficiency, in particular due to the correlations
between transversity angles and slow pion detection efficiencies.
To take these effects into account,
we replace $H_i(\cos\ttr,\cos\tone)$ with distributions of
reconstructed MC events $\calh_i(\cos\ttr,\cos\tone)$,
which are prepared separately for candidates
with and without $\dsp \to D^+\pi^0$ decays
as is done in the branching fraction measurement.
We also introduce effective polarization parameters
$R_i^\prime \equiv
\epsilon_i R_i/(\epsilon_0 R_0 + \epsilon_\| R_\| + \epsilon_\perp R_\perp)$,
where $\epsilon_i$ is a total reconstruction efficiency
for each transversity amplitude.
As a result, the signal probability density function (PDF)
for the fit is defined as
\begin{equation}
\calh_{\rm sig} = \sum_{i} R_i^\prime \calh_i(\cos\ttr,\cos\tone).
\end{equation}

\begin{figure}
\begin{center}
\resizebox{0.4\textwidth}{!}{\includegraphics{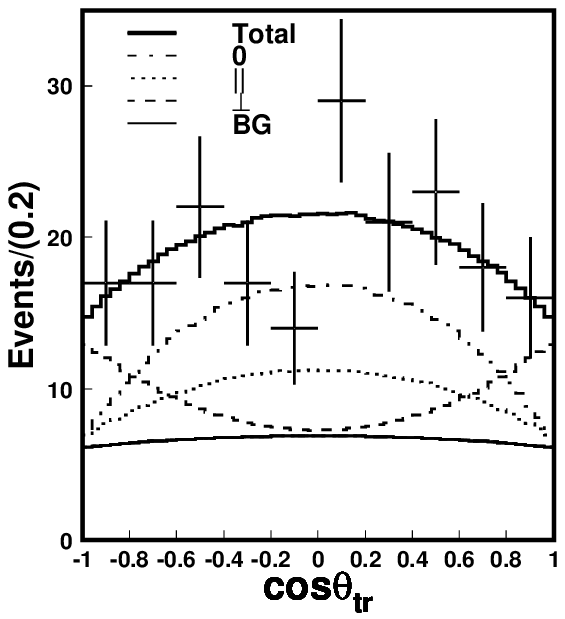}}
\resizebox{0.4\textwidth}{!}{\includegraphics{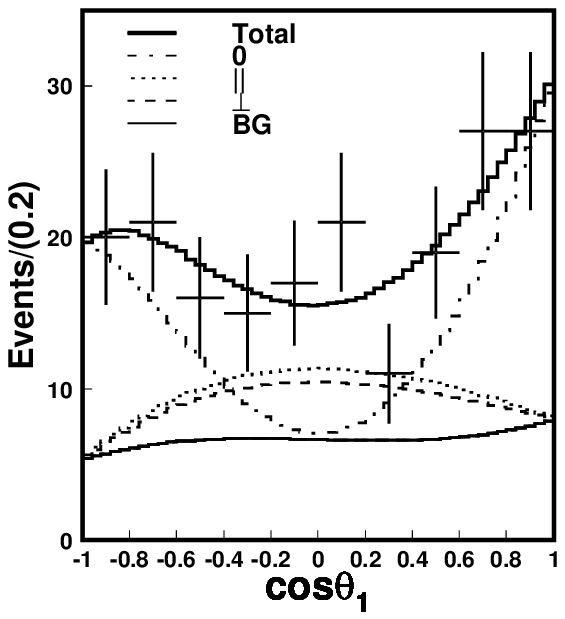}}
\end{center}
\caption{Angular distributions of the $\bztodspdsm$ candidates
in (left) $\cos\ttr$ and (right) $\cos\tone$ projections.
In each figure,
the dot-dashed, dotted and dashed lines correspond to
longitudinal, transverse parallel and transverse perpendicular polarization 
components, respectively.
The thin solid line is the background, and
the thick solid line shows the sum of all contributions.
The asymmetry in the $\cos\tone$ distribution is due to the 
inefficiency for low momentum track reconstruction.}
\label{fig:pol}
\end{figure}

We determine the following likelihood value for each event:
\begin{equation}
{\mathcal L} = \fsig \calh_{\rm sig} + (1-\fsig)\calh_{\rm bg},
\end{equation}
where $\fsig$ is the signal probability calculated on an
event-by-event basis as a function of $\dE$ and $\mb$.
The background PDF $\calh_{\rm bg}$ is determined from
the sideband region ($5.20 $ GeV/$c^2 < \mb < 5.26$ GeV/$c^2$,
$|\de| < 0.2$ GeV).
A fit that maximizes the product of the likelihood values
over all events yields 
\begin{eqnarray}
\rperp &=& \rperpresult, \nonumber \\
\rzero &=& \rzeroresult.
\end{eqnarray}
Figure~\ref{fig:pol} shows the angular distributions with the results of 
the fit.

We study the uncertainties of the following items 
to determine the systematic errors:
background shape parameters, angular resolutions, and slow pion
detection efficiencies. Also included are a possible
fit bias, MC histogram bin size dependence
and misreconstruction effects.
These systematic errors are much smaller than the statistical errors.
%%%%%%%%%%%%%%%%%%%%
%%%%%%%%%%

%%%%%%%%%%%%%%%%%%%%%%%%%%%%%%%%%%%
\section{\boldmath $CP$ Asymmetries}
\label{sec:acp}

We perform an unbinned maximum likelihood fit to the 
three dimensional $\Dt$, 
$\cos\ttr$ and $\cos\tone$ distributions for $\bztodspdsm$ candidates
to measure the $CP$-violation parameters.

The $\bz$ meson decay vertices are reconstructed 
using the $D$ meson trajectory and an IP constraint.
We do not use slow pions from $\dsp$ decays.
We require that at least one $D$ meson has two or more daughter
%tracks with sufficient associated deposits
%in the SVD to precisely measure the $D$ meson trajectory.
tracks with a sufficient number of the SVD hits
to precisely measure the $D$ meson trajectory.
The $\ftag$ vertex determination is the same as for other
$CP$-violation measurements~\cite{bib:CP140PRD_Belle}.

The $b$-flavor of the accompanying $B$ meson is identified
from inclusive properties of particles
that are not associated with the reconstructed $\bz \to \fCP$ 
decay~\cite{bib:CP1_Belle}.
We use two parameters, $\fq$ and $r$, to represent the flavor tagging
information.
The first, $\fq$, is already defined in Eq.~(\ref{eq:psig}).
The parameter $r$ is an event-by-event,
MC-determined flavor-tagging dilution factor
that ranges from $r=0$ for no flavor
discrimination to $r=1$ for unambiguous flavor assignment.
This assignment is used only to sort data into six $r$ intervals.
The wrong tag fractions for the six $r$ intervals, 
$w_l~(l=1,6)$, and differences 
between $\bz$ and $\bzb$ decays, $\dwl$,
are determined from the data;
we use the same values
that were used for the $\sin 2\phi_1$ measurement~\cite{bib:CP140PRD_Belle}.

The signal PDF is given by
\begin{eqnarray}
{\cal P}_{\rm sig}
&=&
\frac{e^{-|\Delta t|/\taub}}{4\taub}
\sum_{i=0,\|,\perp}R_i^\prime {\cal H}_i(\cos\ttr,\cos\tone) \nonumber \\
& &\times\biggl[1-q\Delta w
+q(1-2w)(
\adsds\cos\Delta m\Delta t
+\xi_i\sdsds\sin\Delta m\Delta t)\biggr],
\label{func:cp_pdf}
\end{eqnarray}
where $CP$ parity $\xi_i$ is $+1$ for $i=0$ and $\|$, and
$-1$ for $i=\perp$.
We assume universal $CP$-violation parameters
in Eq.~(\ref{func:cp_pdf}), i.e. 
${\sdsds}_{0} = {\sdsds}_{\|} = {\sdsds}_{\perp}$
and ${\adsds}_{0} = {\adsds}_{\|} = {\adsds}_{\perp}$.
The distribution is convolved with the
proper-time interval resolution function
$R_{\rm sig}(\Dt)$~\cite{bib:CP140PRD_Belle},
which takes into account the finite vertex resolution.

We determine the following likelihood value for the $j$-th event:
\begin{eqnarray}
P_j
&=& (1-\fol)\int \biggl[
\fsig{\mathcal P}_{\rm sig}(\Dt')R_{\rm sig}(\Dt_i-\Dt') \nonumber \\
&+&(1-\fsig){\mathcal P}_{\rm bkg}(\Dt')R_{\rm bkg}(\Dt_i-\Dt')\biggr]
d(\Dt') + \fol P_{\rm ol}(\Dt_i),
\end{eqnarray}
where $P_{\rm ol}(\Dt)$ is a broad Gaussian function that represents
an outlier component~\cite{bib:CP1_Belle} with a small fraction $\fol$.
The $\fsig$ calculation is explained in the previous section.
The PDF for background events, ${\mathcal P}_{\rm bkg}(\Dt)$,
is expressed as a sum of exponential and prompt components, and
is convolved with $R_{\rm bkg}$ that is a sum of two Gaussians. 
All parameters in ${\mathcal P}_{\rm bkg} (\Dt)$
and $R_{\rm bkg}$ are determined by a fit to the $\Dt$ distribution of a 
background-enhanced control sample; 
i.e. events outside of the $\dE$-$\mb$ signal region.
We fix $\tau_\bz$ and $\dmd$ to
their world-average values~\cite{bib:PDG2003}.
The only free parameters in the final fit
are $\cals$ and $\cala$, which are determined by maximizing the
likelihood function
$L = \prod_jP_j(\Dt_j,{\cos\ttr}_j,{\cos\tone}_j;\cals,\cala)$,
where the product is over all events.
The fit yields
\begin{eqnarray}
    \sdsds &=& \sdsdsresult, \nonumber \\
    \adsds &=& \adsdsresult,
\end{eqnarray}
where the first errors are statistical and the second errors
are systematic. 
These results are consistent with the SM expectations
for small penguin contributions.

We define the raw asymmetry in each $\Dt$ bin by
$(N_{q=+1}-N_{q=-1})/(N_{q=+1}+N_{q=-1})$,
where $N_{q=+1(-1)}$ is the number of observed candidates with $q=+1(-1)$.
\begin{figure}
\begin{center}
\resizebox{!}{0.6\textwidth}{\includegraphics{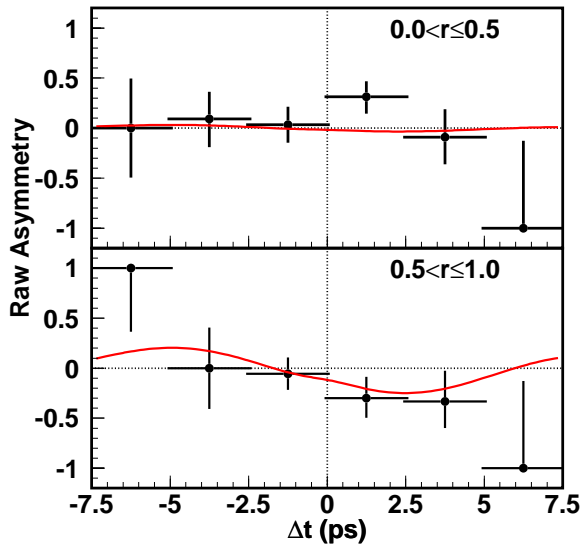}} 
\end{center}
\caption{
Raw $\bztodspdsm$ asymmetry in bins of $\Dt$ for (top) $0 < r \le 0.5$
and $0.5 < r \le 1.0$ (bottom).
The solid curves show the result of the unbinned maximum-likelihood fit.}
\label{fig:asym}
\end{figure}
%%%%%%%%%%
Figure~\ref{fig:asym} shows the raw asymmetries in two regions of 
the flavor-tagging parameter $r$. 
While the numbers of events in the two regions are similar,
the effective tagging efficiency is much larger 
and the background dilution is smaller in the region $0.5 < r \le 1.0$.
Note that these projections onto the $\Delta t$ axis do not take into
account event-by-event information (such as the signal fraction, the
wrong tag fraction and the vertex resolution), which are used in the
unbinned maximum-likelihood fit.

The sources of the systematic errors include
uncertainties in the vertex reconstruction
(0.05 for $\cals$ and 0.03 for $\cala$),
in the flavor tagging
(0.04 for $\cals$ and 0.02 for $\cala$),
in the resolution function
(0.05 for $\cals$ and 0.01 for $\cala$),
in the background fractions
(0.04 for $\cals$ and 0.02 for $\cala$),
in the tag-side interference~\cite{bib:CP140PRD_Belle}
(0.01 for $\cals$ and 0.03 for $\cala$),
and 
in the polarization parameters
(0.06 for $\cals$ and 0.01 for $\cala$).
Other contributions for $\cals$ come from
a possible fit bias (0.04) and 
from uncertainties in $\taubz$ and $\dmd$ (0.02).
We add each contribution in quadrature to obtain the total systematic
uncertainty.

We perform various cross checks. 
A fit to the same sample with 
$\cala$ fixed at zero yields $\cals = -0.69\pm0.56$(stat).
We check with an ensemble of MC pseudo-experiments
that the fit has no sizable bias and the expected statistical
errors are consistent with the measurement.
We also select the following decay modes that
have similar properties to the $\bztodspdsm$ decay:
$B^0\ra D^{*-}D_s^{*+}$, $D^{-}D_s^{*+}$,
$D^{*-}D_s^{+}$, $D^{-}D_s^{+}$, and
$B^+\ra \dzerobstar D_s^{*+}$, $\dzerob D_s^{*+}$,
$\dzerobstar D_s^{+}$ and $\dzerob D_s^{+}$.
Fits to the control samples yield
$\cals[B^0\ra D^{(*)}D_s^{(*)}] = -0.12 \pm 0.08$,
$\cala[B^0\ra D^{(*)}D_s^{(*)}] = +0.02 \pm 0.05$,
$\cals[B^+\ra D^{(*)}D_s^{(*)}] = -0.10 \pm 0.07$, and
$\cala[B^+\ra D^{(*)}D_s^{(*)}] = -0.001 \pm 0.050$, where
errors are statistical only.
All results are consistent with zero.
We also measure the $B$ meson lifetime
using $\bztodspdsm$ candidates as well as 
the control samples.
All results are consistent with the present world-average values.
A fit to the $\Delta t$ distribution of the $\bztodspdsm$ without using
polarization angle information yields
$\sdsds = -0.57 \pm 0.45$,
$\adsds = -0.29 \pm 0.26$;
this result suggests that the $CP$-odd component is small,
supporting our polarization measurement.

Although the statistics are not sufficient to provide tight constraints,
we also consider polarization-dependent values for $\cals$ and $\cala$,
which may arise from possible differences in 
the contributions of the penguin diagrams. 
We assume that the $CP$ asymmetries for the $CP$-odd component
are consistent with the SM expectations, and fix
${\sdsds}_{\perp}$ at the world-average value of $\sinbb$~\cite{bib:PDG2003}
and ${\adsds}_{\perp}$ at zero.
A fit with this assumption yields
$\sdsds = -0.72 \pm 0.50$ and
$\adsds = -0.42 \pm 0.30$
for the $CP$-even component, also consistent with the SM expectations.

%%%%%%%%%%%%%%%%%%%%%%%%%%%%%%
\section{Conclusion}
\label{sec:conclusion}

In summary, we have performed measurements of 
the branching fraction, the polarization parameters and the
$CP$-violation parameters for $\bztodspdsm$ decays.
The results are
\begin{eqnarray}
\calb(\bztodspdsm) &=& \brdsdsresult, \nonumber \\
\rperp &=& \rperpresult, \nonumber \\
\rzero &=& \rzeroresult, \nonumber \\
\sdsds &=& \sdsdsresult, \nonumber \\
\adsds &=& \adsdsresult.
\end{eqnarray}
The polarization parameters and $CP$-violation parameters
are consistent with the SM expectations
and theoretical predictions for small penguin 
contributions~\cite{bib:sm_prediction}.

%%%%%%%%%%%%%%%%%%%%%%%%%%%%%%%%%%%
\section*{Acknowledgments}
\label{sec:acknowledgment}

  We thank the KEKB group for the excellent
  operation of the accelerator, the KEK Cryogenics
  group for the efficient operation of the solenoid,
  and the KEK computer group and the NII for valuable computing and
  Super-SINET network support.  We acknowledge support from
  MEXT and JSPS (Japan); ARC and DEST (Australia); NSFC (contract
  No.~10175071, China); DST (India); the BK21 program of MOEHRD and the
  CHEP SRC program of KOSEF (Korea); KBN (contract No.~2P03B 01324,
  Poland); MIST (Russia); MESS (Slovenia); NSC and MOE (Taiwan); and DOE
  (USA).


\begin{thebibliography}{999}
\bibitem{bib:CC}
Throughout this paper,
the inclusion of the charge conjugate decay mode is implied
unless otherwise stated.
\bibitem{bib:sm_prediction}
 If penguin contributions are small, the theoretical predictions 
 within the SM are ${\mathcal A}\simeq 0$
 and ${\mathcal S}\simeq -\sin 2\phi_1$, where
 $\sin 2\phi_1$ is measured in $\btoccs$ 
 transitions to be $0.731\pm 0.056$~\cite{bib:PDG2003}.
\end{thebibliography}
\end{document}